\documentstyle[12pt,aaspp4,psfig]{article}

\lefthead{Matsubara} \righthead{STOCHASTICITY OF BIAS AND NONLOCALITY
OF GALAXY FORMATION}

\begin{document}
\newcommand{\gsim}{\mbox{\raisebox{-1.0ex}{$\stackrel{\textstyle >}
{\textstyle \sim}$ }}}
\newcommand{\lsim}{\mbox{\raisebox{-1.0ex}{$\stackrel{\textstyle <}
{\textstyle \sim}$ }}}

\newcommand{\gtsima}{$\; \buildrel > \over \sim \;$}
\newcommand{\ltsima}{$\; \buildrel < \over \sim \;$}
\newcommand{\simgt}{\lower.5ex\hbox{\gtsima}}
\newcommand{\simlt}{\lower.5ex\hbox{\ltsima}}
\newcommand{\himpc}{{\hbox {$h^{-1}$}{\rm Mpc}} }
\newcommand{\bfk}{{\mbox{\boldmath $k$}}}
\newcommand{\bfq}{{\mbox{\boldmath $q$}}}
\newcommand{\bfr}{{\mbox{\boldmath $r$}}}
\newcommand{\bfx}{{\mbox{\boldmath $x$}}}
\newcommand{\bfy}{{\mbox{\boldmath $y$}}}
\newcommand{\bfz}{{\mbox{\boldmath $z$}}}
\newcommand{\bfv}{{\mbox{\boldmath $v$}}}
\newcommand{\sbfk}{{\mbox{\scriptsize\boldmath $k$}}}
\newcommand{\sbfx}{{\mbox{\scriptsize\boldmath $x$}}}
\newcommand{\bfpsi}{{\mbox{\boldmath $\psi$}}}
\newcommand{\bfPsi}{{\mbox{\boldmath $\Psi$}}}
\newcommand{\mPsi}{{\mit\Psi}}
\newcommand{\delg}{\delta_{\rm g}}
\newcommand{\delm}{\delta_{\rm m}}
\newcommand{\deli}{\delta_{\rm i}}
\newcommand{\dell}{\delta_{\rm L}}
\newcommand{\tdell}{\widetilde{\delta}_{\rm L}}
\newcommand{\dells}{\delta_{{\rm L}s}}
\newcommand{\xil}{\xi_{\rm L}}
\newcommand{\sigg}{\sigma_{\rm g}}
\newcommand{\sigm}{\sigma_{\rm m}}
\newcommand{\sigls}{\sigma_{{\rm L}s}}
\newcommand{\rhog}{\rho_{\rm g}}
\newcommand{\rhom}{\rho_{\rm m}}
\def\pp{\par\parshape 2 0truecm 15.5truecm 1truecm 14.5truecm\noindent}
\renewcommand{\theequation}{\mbox{\rm 
{\arabic{section}.\arabic{equation}}}}


\title{STOCHASTICITY OF BIAS AND NONLOCALITY OF GALAXY FORMATION:
LINEAR SCALES}

\author{Takahiko Matsubara}

\affil{Department of Physics and Astronomy, The Johns Hopkins
	University, 3400 N.Charles Street, Baltimore, MD 21218; and\\
	Department of Physics, The University of Tokyo, Hongo 7-3-1,
	Tokyo 113-0033, Japan; and\\ Research Center for the Early
	Universe, Faculty of Science, The University of Tokyo, Tokyo
	113-0033, Japan.}

\begin{abstract}
If one wants to represent the galaxy number density at some point in
terms of only the mass density at the same point, there appears the
stochasticity in such a relation, which is referred to as ``stochastic
bias''. The stochasticity is there because the galaxy number density
is not merely a local function of a mass density field, but it is a
nonlocal functional, instead. Thus, the phenomenological stochasticity
of the bias should be accounted for by nonlocal features of galaxy
formation processes. Based on mathematical arguments, we show that
there are simple relations between biasing and nonlocality on linear
scales of density fluctuations, and that the stochasticity in Fourier
space does not exist on linear scales under a certain condition, even
if the galaxy formation itself is a complex nonlinear and nonlocal
precess. The stochasticity in real space, however, arise from the
scale-dependence of bias parameter, $b$. As examples, we derive the
stochastic bias parameters of simple nonlocal models of galaxy
formation, i.e., the local Lagrangian bias models, the cooperative
model, and the peak model. We show that the stochasticity in real
space is also weak, except on the scales of nonlocality of the galaxy
formation. Therefore, we do not have to worry too much about the
stochasticity on linear scales, especially in Fourier space, even if
we do not know the details of galaxy formation process.
\end{abstract}



\section{INTRODUCTION}
\setcounter{equation}{0}

Probing the statistical properties of the large-scale structure of the
universe has a great importance in studying the origin of our
universe. Recent galaxy surveys have been revealing the statistical
properties of distribution of galaxies on very large scales. The
number density of galaxies are not necessarily proportional to the
density of mass, and this ambiguity is known as galaxy biasing problem
(Kaiser 1984; Davis et al. 1985; Bardeen et al. 1986). Since galaxies
of different types have different clustering properties (e.g.,
Dressler 1980; Davis \& Geller 1976; Giovanelli, Haynes, \& Chincarini
1986; Santiago \& Strauss 1992; Loveday et al. 1996; Hermit et al.
1996; Guzzo et al. 1997), not all types of galaxies can simultaneously
be unbiased tracers of mass. This ambiguity is undesirable in
extracting cosmological information from the data of galaxy
distribution.

The simplest model for the galaxy bias is the {\em local}, {\em
linear} bias. In this simple model, the number density field of galaxy
with a fixed smoothing scale $R$ is assumed to be proportional to
density field of mass with a same smoothing scale:
\begin{eqnarray}
   \delg (\bfx;R) = b\, \delm (\bfx;R),
\label{eq0}
\end{eqnarray}
where $\delg$ and $\delm$ are density contrast of galaxy and of mass,
respectively, with a fixed smoothing length $R$. This model is viable
if (a) $\delg(\bfx;R)$ is dependent only on $\delm(\bfx;R)$ and (b)
density contrast of mass is sufficiently small. The latter condition
(b) is achieved by considering large scales on which density
fluctuations are small enough so that only linear term becomes
significant. The linear coefficient $b$ is called as bias parameter.
The bias parameter is often assumed to be a constant, although it can
depend on $R$ in general. The effect of nonlinearity should be taken
into account when we are interested in nonlinear scales. On weakly
nonlinear scale, this affect the estimation of higher order statistics
(e.g., Fry \& Gazta\~naga 1993).

Although the latter condition is reasonably considered to be valid if
we are interested in linear scales, the former condition (a) is not
trivial so far. The non-triviality of condition (a) leads us to a
concept of stochastic bias which is recently argued (Dekel \& Lahav
1998; Pen 1998; Tegmark \& Peebles 1998; Tegmark \& Blomley 1998;
Taruya, Koyama \& Soda 1998; Taruya \& Soda 1998; Blanton et al.
1998). In the stochastic biasing scheme, $\delg(\bfx;R)$ is not
supposed to be determined solely by $\delm(\bfx;R)$, but the scatter
in $\delg$-$\delm$ relation is taken into account. In linear regime in
which the density contrast of mass $\delm$ is small enough, and is
approximated by random Gaussian field, the two-point statistics fully
characterize the statistics of the scatter. In literatures, the bias
parameter $b$ and the dimensionless cross correlation $r$ are used to
characterize the linear stochastic biasing scheme:
\begin{eqnarray}
   b(R) = \sqrt{\frac{\langle \delg(\bfx;R)^{\,2} \rangle}
                     {\langle \delm(\bfx;R)^{\,2} \rangle}},
   \quad
   r(R) =  \frac{\langle \delm(\bfx;R) \delg(\bfx;R) \rangle}
                {\sigm(R) \sigg(R)},
\label{eq1}
\end{eqnarray}
where $\sigm = \langle\delm^{\,2}\rangle^{1/2}$ and $\sigg =
\langle\delg^2\rangle^{1/2} = b \sigm$ are rms density fluctuations of
mass and galaxies, respectively.

The bias parameter $b$ in the equation (\ref{eq1}) is a generalization
of the bias parameter in linear deterministic biasing scheme of
equation (\ref{eq0}). In deterministic case, the cross correlation $r$
is always unity. {}From Schwarz inequality, $r$ cannot exceed unity
and $r=1$ means that biasing is deterministic, $\delg \propto \delm$.
Thus, the deviation from $r=1$ measures the stochasticity. If the
smoothing scale $R$ is large enough so that $\delm$ and $\delg$ are
considered as bivariate Gaussian field, these three parameters $b$,
$r$ and $\sigm$ contain all the statistical information about the
stochastic biasing.

In literatures, these parameters are sometimes considered as free
parameters to be determined by observation. However, if we could know
the process of galaxy formation in detail, the parameters $b$ and $r$
would be derived from some fundamental physical processes. This is
because the bias and its stochasticity come from our ignorance of the
galaxy formation. Therefore, there should be some theoretical
constraint between $b$ and $r$. At first sight, it seems difficult to
find any constraint as we do not exactly know the process of galaxy
formation. This is true especially on small scales where the nonlinear
and nonlocal characters of galaxy formation plays an important role on
the statistics of galaxy distributions. This problem is one of the
most important issues in astrophysics and much numerical and
analytical work is needed (e.g., Rees \& Ostriker 1977; White \& Frenk
1991; Cen \& Ostriker 1992; Mo \& White 1996). However, on large
scales, such undesirable characters can be expected to be small.
Scherrer \& Weinberg (1998), based on the local biasing scheme, showed
that the stochasticity actually vanishes on large scales and galaxy
autocorrelation function behaves exactly as in deterministic biasing
scheme. Dekel \& Lahav (1999) also imply the same property based on a
specific simple model.

In this paper, based on a general nonlocal method, we show that
stochasticity in Fourier space asymptotically vanishes on linear
scales under a certain condition, explicitly deriving the relation
between the stochastic parameters $b$, $r$ and the nonlinear, nonlocal
functional form of galaxy formation. We will show the first
coefficient of generalized Wiener-Hermite functional, which is defined
below, of the nonlocal, nonlinear relations of galaxy formation will
contribute to the galaxy statistics on large scales, if that
coefficient does not vanish.

In the derivation, we can use the technique developed in Matsubara
(1995), in which the diagrammatic methods for the calculation of
general nonlocal biasing are introduced. The diagrammatics are useful
especially when the non-Gaussianity and/or higher-order correlations
are interested in. In this paper, however, we derive the result
without employing diagrammatics for self-consistency and for the
simplicity of the problem. The diagrammatics make it easier to
generalize the present results to higher-order statistics.

In \S 2, we revisit the mathematical methods for nonlinear, and
nonlocal bias, which have been developed by Matsubara (1995). Then we
derive the relation between the stochastic bias parameters and the
nonlocality of biasing. In \S 3, we examine three types of biasing
schemes, i.e., the local Lagrangian bias models, the cooperative
model, and the peak model, according to the result of \S 2. In \S 4,
we discuss the results and present the conclusions.

\section{NONLOCAL BIAS AND STOCHASTICITY}
\setcounter{equation}{0}

\subsection{Mathematical Formulation of Nonlocal Bias}

In the present paradigm, the distribution of galaxies is determined by
initial density fluctuations. Whether a galaxy has formed at some
place or not should be fully determined by the initial fluctuations.
In this sense, the galaxy formation is deterministic, although it
nonlinearly, nonlocally, and possibly chaotically depends on initial
density fluctuations. In the stochastic biasing scheme, this complex
features of nonlinearity and nonlocality of the galaxy formation are
expressed by phenomenological scatter of local $\delg$--$\delm$ relation.
Thus, in principle, stochasticity can be determined by nonlinear and
nonlocal deterministic processes of galaxy formation.

Since both the present mass density field $\rhom$ and the density
field of galaxies $\rhog$ are determined by initial density
fluctuations $\deli$, they are expressed by functionals $F_{\rm m}$
and $F_{\rm g}$. In the following, instead of the initial fluctuations
$\deli$, we will alternatively use linearly extrapolated density
fluctuations $\dell = D\deli$, where $D$ is a linear growth rate. The
variable $\dell$ is simply a linear extrapolation of the evolution of
density contrast, regardless of whether or not the small scale
fluctuations are actually in linear regime. The introduction of the
linearly extrapolated field is just for convenience and it simply
represents the initial density fluctuations. In this notation, one can
write the relations as
\begin{eqnarray}
   \rhom(\bfx) = F_{\rm m}([\dell],\bfx),\qquad
   \rhog(\bfx) = F_{\rm g}([\dell],\bfx),
   \label{eq2}
\end{eqnarray}
where we introduce the notation $[\dell]$ which means that $F_{\rm m}$
and $F_{\rm g}$ are nonlocal (and, of course, nonlinear) functionals
of $\dell$. These functionals also depend on the position, $\bfx$. It
is sometimes more convenient to use the density contrast of $\rhom$
and $\rhog$:
\begin{eqnarray}
   \delm(\bfx) = 
   \frac{F_{\rm m}([\dell],\bfx)}
      {\langle F_{\rm m}([\dell],\bfx)\rangle} - 1,\qquad
   \delg(\bfx) = 
   \frac{F_{\rm g}([\dell],\bfx)}
      {\langle F_{\rm g}([\dell],\bfx)\rangle} - 1.
   \label{eq2-1}
\end{eqnarray}

One may want to expand these functional by Taylor series of $\dell$,
as in usual perturbative approach. However, this approach would be
useful only if the nonlinearities on {\em all} scales were small,
and/or galaxy formation process were linear or quasi-linear. In
reality, however, the galaxy formation process is a highly nonlinear
process, and it depends on the nonlinear behavior of mass density
field in a very complex way. Thus, we cannot use Taylor expansion in
our treatment below. Alternatively, we will find below that the
expansion of these functional by orthogonal functionals, i.e.,
generalized Wiener-Hermite functionals, is useful for our purpose. The
$n$-th order generalized Wiener-Hermite functional ${\cal H}^{(n)}$,
which is introduced in Matsubara (1995) for the first time, is defined
by
\begin{eqnarray}
&&
   {\cal H}^{(n)}(\bfx_1,\ldots,\bfx_n) 
   = \exp\left[ \frac12\int d^3x d^3y \dell(\bfx)
                \xil^{-1}(\bfx,\bfy) \dell(\bfy) \right]
\nonumber \\
&&\qquad\qquad
   \times\,\frac{(-1)^n \delta^n}{\delta \dell(\bfx_1) 
                            \cdots \delta \dell(\bfx_n)}
     \exp\left[ -\frac12\int d^3x d^3y \dell(\bfx)
                \xil^{-1}(\bfx,\bfy) \dell(\bfy) \right],
   \label{eq3}
\end{eqnarray}
where $\delta/\delta\dell$ is a functional derivative with regard to
$\dell$, and $\xil$ is a two-point correlation function of linear
density field, $\xil(\bfx,\bfy) =
\langle\dell(\bfx)\dell(\bfy)\rangle$. The inverse of correlation
function, $\xil^{-1}$, is defined formally by
\begin{equation}
   \int d^3y \xil^{-1}(\bfx,\bfy) \xil(\bfy,\bfz)
   = \delta^3(\bfx - \bfz).
   \label{eq4}
\end{equation}

The original Wiener-Hermite functional is the special case when the
correlation function is identical to Dirac delta function,
$\xil(\bfx,\bfy) = \delta^3(\bfx-\bfy)$. If $\dell$ has only finite
degrees of freedom instead of functional of space, the generalized
Wiener-Hermite functionals reduce to generalized Wiener-Hermite
polynomials which are seen in the literature (Appel \& F\'eriet 1926).
For the application to modern cosmology, the generalized
Wiener-Hermite functionals are more appropriate than the others.

We expand the functionals $F_{\rm m}$ and $F_{\rm g}$ by these
functionals. It is useful to define the following functional which is
an infinite dimensional linear combination of ${\cal H}^{(n)}$:
\begin{equation}
   {\cal H}_{(n)}(\bfx_1,\ldots,\bfx_n)
   = \int d^3y_1 \cdots d^3y_n \xil(\bfx_1,\bfy_1) 
     \cdots \xil(\bfx_n,\bfy_n) {\cal H}^{(n)}(\bfy_1,\ldots,\bfy_n),
   \label{eq5}
\end{equation}
which we also call as generalized Wiener-Hermite functional. The
explicit form of the first three generalized Wiener-Hermite
functionals are
\begin{eqnarray}
&&   {\cal H}_{(0)} = 1,
\\&& {\cal H}_{(1)}(\bfx) = \dell(\bfx),
\\&& {\cal H}_{(2)}(\bfx_1,\bfx_2) =
     \dell(\bfx_1)\dell(\bfx_2) - \xil(\bfx_1-\bfx_2),
\\&& {\cal H}_{(3)}(\bfx_1,\bfx_2,\bfx_3) =
     \dell(\bfx_1)\dell(\bfx_2)\dell(\bfx_3)
\\&&\qquad -\,
     [\xil(\bfx_1-\bfx_2)\dell(\bfx_3) +
      \xil(\bfx_2-\bfx_3)\dell(\bfx_1) +
      \xil(\bfx_3-\bfx_1)\dell(\bfx_2)].
\end{eqnarray}

The generalized Wiener-Hermite functionals have the orthogonality
relation:
\begin{eqnarray}
   &{}&\left\langle {\cal H}_{(n)}(\bfx_1,\ldots,\bfx_n)
                {\cal H}^{(m)}(\bfy_1,\ldots,\bfy_m)
   \right\rangle
  \nonumber \\
   &{}& \qquad = \delta_{nm}\left[
       \delta^3(\bfx_1-\bfy_1)\cdots\delta^3(\bfx_n-\bfy_n)
       + {\rm sym.(}\bfy_1,\ldots,\bfy_n{\rm )}\right],
   \label{eq6}
\end{eqnarray}
where ${\rm sym.(...)}$ denote the terms of symmetric permutation of
the preceding term with respect to the indices. In this case, there
are $n!$ terms in total in the right hand side of equation
(\ref{eq6}). The Gaussian averaging $\langle\cdots\rangle$ denote the
averaging by linear density field $\dell$ by the random Gaussian
probability functional ${\cal P}[\dell]$:
\begin{eqnarray}
   \langle\cdots\rangle =
   \int [d\dell] \cdots {\cal P}[\dell],
   \label{eq7}
\end{eqnarray}
where
\begin{equation}
   {\cal P}[\dell] = 
	{\cal N} \exp \left[ -\frac12 \int d^3x d^3y
      \dell(\bfx) \xil^{-1}(\bfx,\bfy) \dell(\bfy) \right].
   \label{eq8}
\end{equation}
The formal normalization constant ${\cal N}$ is given by
\begin{equation}
   {\cal N} = \left\{ 
      \int [d\dell] \exp \left[ -\frac12 \int d^3x d^3y
      \dell(\bfx) \xil^{-1}(\bfx,\bfy) \dell(\bfy) \right]
         \right\}^{-1}.
   \label{eq9}
\end{equation}
Because the degrees of freedom is infinite, ${\cal N}^{\,-1}$ formally
diverges, but the proper regularization is always possible by
discretizing the three dimensional continuum space. One can prove the
orthogonality equation (\ref{eq6}), just generalizing the proof of
orthogonality of simple Hermite polynomial, which is well known.

We consider the functionals of equation (\ref{eq5}) as base
functionals for the expansion of the functional of mass density field,
and of galaxy density field:
\begin{equation}
   \delta_{\rm A}(\bfr,[\dell]) = \sum_{n=1}^\infty \frac1{n!}
      \int d^3x_1 \cdots d^3x_n
      K_{\rm A}^{(n)}(\bfr-\bfx_1,\cdots,\bfr-\bfx_n)
      {\cal H}_{(n)}(\bfx_1,\ldots,\bfx_n),
   \label{eq10}
\end{equation}
where A = m or g. The reason why $n=0$ term is not appeared in the
above expression is that $\langle\delta_{\rm A}\rangle = 0$ and
$\langle{\cal H}_{(n)}\rangle = 0$ for $n>0$ [set $m=0$ in equation
(\ref{eq6})]. This expansion is complete because the kernel $K^{(n)}$
is uniquely given by
\begin{eqnarray}
   K_{\rm A}^{(n)}(\bfr-\bfx_1,\ldots,\bfr-\bfx_n)
   &=& \left\langle {\cal H}^{(n)}(\bfx_1,\ldots,\bfx_n)
      \delta_{\rm A}(\bfr,[\dell])\right\rangle
   \label{eq11} \\
   &=& \left\langle \frac{\delta^n \delta_{\rm A}(\bfr,[\dell])}
       {\delta\dell(\bfx_1)\cdots \delta\dell(\bfx_n)}
      \right\rangle.
   \label{eq12}
\end{eqnarray}

According to the expansion (\ref{eq10}) and the orthogonality relation
(\ref{eq6}), the two-point auto-correlation function of matter
$\xi_{\rm mm}$ and of galaxies $\xi_{\rm gg}$ and the cross
correlation function $\xi_{\rm mg}$ are given by
\begin{eqnarray}
   \xi_{\rm AB}(\bfr) &=&
   \langle\delta_{\rm A}(\bfr_1)\delta_{\rm B}(\bfr_2)\rangle
\nonumber\\
   &=& \sum_{n=1}^\infty \frac{1}{n!}
   \int d^3x_1\cdots d^3x_n d^3y_1\cdots d^3y_n
   K_{\rm A}^{(n)}(\bfx_1,\ldots,\bfx_n)
   K_{\rm B}^{(n)}(\bfy_1,\ldots,\bfy_n)
\nonumber\\
&&\qquad\qquad \times\,
   \xil(\bfr + \bfx_1 + \bfy_1) \xil(\bfr + \bfx_2 + \bfy_2) \cdots
   \xil(\bfr + \bfx_n + \bfy_n),
   \label{eq14}
\end{eqnarray}
where A, B = m, g, and $\bfr_1 - \bfr_2 = \bfr$. Note that this
expansion is valid as long as $\dell$ is a random Gaussian field. If
the initial density field is non-Gaussian, there are additional terms
in equation (\ref{eq14}) which depend on initial higher-order
correlation functions [see Matsubara (1995) for detail].

Both the nonlocal kernels $K_{\rm m}$ and $K_{\rm g}$ do not depend on
$\bfr$. Thus, on large scales with respect to the separation $\bfr$,
we can approximate equation (\ref{eq14}) by only considering lower
order terms of $\xil$, provided that the kernel $K_{\rm A}^{(n)}$ does
not have broad profile.

If the kernel falls off slowly on large scales, we can not truncate
the expansion (\ref{eq14}). In the rest of this paper, we assume the lowest
order term in the expansion (\ref{eq14}) actually dominates the
higher-order terms. Before proceeding to the analysis of the lowest
order approximation, we consider the cases in which this assumption
breaks down. Imagine, for example, that $\xi_{\rm L}(r) \sim
r^{-\gamma}$, and $K^{(n)}_{\rm A}(\bfx_1,\ldots,\bfx_n) \sim
(x_1\cdots x_n)^{-s_{\rm A}}$ on large scales. Then the $n$-th order
term in the integral in expansion (\ref{eq14}) is approximately given
by
\begin{eqnarray}
&&
   \int 
   \frac{d^3x_1\cdots d^3x_n d^3y_1\cdots d^3y_n}
	{(x_1\cdots x_n)^{s_{\rm A}}
   	 (y_1\cdots y_n)^{s_{\rm B}}
	 |\bfr + \bfx_1 + \bfy_1|^\gamma \cdots
	 |\bfr + \bfx_n + \bfy_n|^\gamma
	} 
\nonumber\\
&& \qquad
   = \left(
       \int
       \frac{d^3x d^3y}{x^{s_{\rm A}} y^{s_{\rm B}}
             |\bfr + \bfx + \bfy|^\gamma}
     \right)^n
   \propto \left(r^{6 - s_{\rm A} - s_{\rm B} - \gamma}\right)^n.
   \label{eq14.5}
\end{eqnarray}
The last expression is derived by the fact that the integral is the
form of convolution and the (3D) Fourier transform of $k^n$ is
proportional to $r^{-n-3}$ (Peebles 1980). As seen by this expression,
$6 - s_{\rm A} - s_{\rm B} - \gamma$ should be negative to ensure the
equation (\ref{eq14.5}) actually falls off on large scales. For
example, if $K^{(2)}_{\rm g}$ falls off as $x_1^{-2} x_2^{-2}$, and
$\gamma \leq 2$, then equation (\ref{eq14.5}) does not fall off
anymore. On the other hand, if $s_{\rm A,B} \geq 3$, then equation
(\ref{eq14.5}) falls off rapidly enough so that the higher order terms
can be neglected. In the latter case, the spatial integration of
$K^{(n)}$ is finite. It is a natural assumption that the spatial
integration of $K^{(n)}$ is finite. If it is infinite, as seen from
equation (\ref{eq10}), the information of density fluctuations of
infinitely distant places affect the galaxy formation as much as, or
more than the fluctuations of nearer places, which is unlikely in
reality. We assume $s_{\rm A,B} \geq 3$ in the rest of this paper. In
other words, we assume that Fourier transform of higher-order kernels
$\widetilde{K}^{(n)}_{\rm A}(\bfk_1,\ldots,\bfk_n)$ are finite in the
limit $k_1,\ldots,k_n \rightarrow 0$.

Note that the expansion (\ref{eq14}) is essentially different from
usual perturbative approach by Taylor expansion of density contrast
itself. Instead, we employ the orthogonal expansion for galaxy and
mass density fields, and the resulting expression, (\ref{eq14}), can
be interpreted as an asymptotic expansion by correlation function,
$\xil$. Thus, we only assume the smallness of correlation function
$\xil$ on large scales, whether the density contrast on small scale is
large or not.

\subsection{A Sample Calculation of Kernels}

Since our argument in this section is extremely formal, it would be
instructive to calculate an example. Consider a model in which $\delg
= \delm + \delm^2$ and $\delm$ is calculated up to second order in
perturbation theory [e.g., Fry (1984)]:
\begin{eqnarray}
   \delm(\bfr) =
   \dell(\bfr) +
   \int d^3x_1 d^3x_2 R(\bfr-\bfx_1, \bfr-\bfx_2)
   \dell(\bfx_1) \dell(\bfx_2),
   \label{eq-a1}
\end{eqnarray}
where
\begin{eqnarray}
   R(\bfx_1, \bfx_2) =
   \int \frac{d^3k_1}{(2\pi)^3} \frac{d^3k_2}{(2\pi)^3}
   \left[
      \frac57 +
      \frac{\bfk_1\cdot\bfk_2}{2 k_1 k_2}
      \left(\frac{k_1}{k_2} + \frac{k_2}{k_1} \right) +
      \frac27
      \left(\frac{\bfk_1\cdot\bfk_2}{k_1 k_2}\right)^2
   \right]
   e^{i \bfk_1 \cdot \bfx_1 + i \bfk_2 \cdot \bfx_2},
   \label{eq-a2}
\end{eqnarray}
and we assume the Einstein-de Sitter Universe for simplicity. Then,
according to equation (\ref{eq12}), we can calculate the kernels as
\begin{eqnarray}
&&
   K^{(1)}_{\rm m}(\bfx) = \delta^3(\bfx),
\\&&
   K^{(2)}_{\rm m}(\bfx_1,\bfx_2) = 
   2 R(\bfx_1,\bfx_2),
\\&&
   K^{(1)}_{\rm g}(\bfx) =
   \delta^3(\bfx) + 
   2 \int d^3y R(\bfx,\bfy)\xil(\bfy)
\\&& \qquad\qquad\qquad +\, 
   2\, \delta^3(\bfx) \int d^3y_1 d^3y_2
   R(\bfy_1,\bfy_2) \xil(\bfy_1 - \bfy_2),
\\&&
   K^{(2)}_{\rm g}(\bfx_1,\bfx_2) =
   2\delta^3(\bfx_1)\delta^3(\bfx_2) + 
   2 R(\bfx_1,\bfx_2)
\\&& \qquad\qquad\qquad +\, 
   4 R(\bfx_1,\bfx_2)
      \int d^3y_1 d^3y_2 R(\bfy_1,\bfy_2)\xil(\bfy_1-\bfy_2)
\\&& \qquad\qquad\qquad +\, 
   8 \int d^3y_1 d^3y_2 R(\bfx_1,\bfy_1) R(\bfx_2,\bfy_2)
      \xil(\bfy_1-\bfy_2).
\end{eqnarray}
Since $R(\bfx_1,\bfx_2)$ drops off as $|\bfx_1|^{-3}|\bfx_2|^{-3}$ on
large scales, the discussion at the end of the previous subsection
suggests that $n$-th order term in equation (\ref{eq14}) actually
drops off as $(\xil)^n$ on large scales.

\subsection{Stochastic Parameters in Linear Regime}

In the following, we are interested in large scales and consider only
the lowest order approximation of equation (\ref{eq14}):
\begin{eqnarray}
   \xi_{\rm AB}(\bfr) =
   \int d^3x d^3y
   K_{\rm A}^{(1)}(\bfx)
   K_{\rm B}^{(1)}(\bfy)
   \xil(\bfr + \bfx + \bfy),
   \label{eq15}
\end{eqnarray}
assuming that this term does not vanish and that the higher-order
terms are negligible. Since this expression has the form of
convolution, it becomes just products in Fourier space. From
statistical isotropy, the Fourier transform of $K^{(1)}$, denoted as
$\widetilde{K}^{(1)}$, is a function of the absolute value of the wave
vector $k = |\bfk|$:
\begin{eqnarray}
   P_{\rm AB}(k) =
   \widetilde{K}_{\rm A}^{(1)}(k) \widetilde{K}_{\rm B}^{(1)}(k)
   P_{\rm L}(k),
   \label{eq16}
\end{eqnarray}
where $P_{\rm L}(k)$ is the linear power spectrum\footnote{If we
consider $P_{\rm AB}(k)$ as the true spectrum, there is possibly a
constant term in addition to the above equation, which comes from the
small scale inaccuracy of expression (\ref{eq15}) [see Scherrer \&
Weinberg (1998); Dekel \& Lahav (1998)]. However, in the following, we
consider $P_{\rm AB}(k)$ as a merely mathematical quantity which
represents just the Fourier transform of $\xi_{\rm AB}(r)$ of the
equation (\ref{eq15}).}.

The linear-scale power spectrum of mass is simply given by $P_{\rm
mm}(k) = P_{\rm L}(k)$. This means $\widetilde{K}_{\rm m}^{(1)}(k) =
1$ on linear scales. Thus, denoting $b_{\rm F}(k) = \widetilde{K}_{\rm
g}^{(1)}(k)$,
\begin{eqnarray}
&&
   P_{\rm mm}(k) = P_{\rm L}(k),
\label{eq17}\\&&
   P_{\rm mg}(k) = b_{\rm F}(k) P_{\rm L}(k),
\label{eq18}\\&&
   P_{\rm gg}(k) = b_{\rm F}^{\,2}(k) P_{\rm L}(k).
\label{eq19}
\end{eqnarray}
These equations are valid as long as $\widetilde{K}_{\rm g}^{(1)}(k)
\ne 0$. If $\widetilde{K}_{\rm g}^{(1)}(k) = 0$, equation (\ref{eq16})
vanishes and is no longer the lowest order in the expansion of
equation (\ref{eq14}). In such case, we should consider the
higher-order terms.

In equations (\ref{eq17})--(\ref{eq19}), $b_{\rm F}(k)$ can be
identified to the linear bias parameter in Fourier space, and is given
by, from Fourier transform of equation (\ref{eq12}),
\begin{eqnarray}
   b_{\rm F}(k) =
   \int d^3x e^{i\sbfk\cdot\sbfx}
   \left\langle 
      \frac{\delta\delg({\bf 0})}{\delta\dell(\bfx)}
   \right\rangle =
   \left\langle
      \frac{\delta\delg({\bf 0})}{\delta\tdell(\bfk)}
   \right\rangle,
\label{eq20}
\end{eqnarray}
where $\tdell(\bfk)$ is a Fourier transform of linear density
fluctuations, $\dell(\bfx)$. 

These simple equations (\ref{eq17})--(\ref{eq20}) are the primary
results of this paper. These equations show, as long as
$\widetilde{K}_{\rm g}^{(1)}(k) \ne 0$, that there are no residual
cross correlation in Fourier space in linear regime (except for the
constant term which comes from the small scale behavior of correlation
function) and that the bias parameter in Fourier space is $b_{\rm
F}(k)$ which is scale-dependent and is related to nonlinearity and
nonlocality of galaxy formation through equation (\ref{eq20}). This
means that the Fourier-mode stochasticity which arise from the
nonlinearity and nonlocality of the galaxy formation vanishes in
linear regime. Thus, the cross correlation $r(k)$ in Fourier space
should approach to unity in large-scale limit, as long as the galaxy
formation is such that $\widetilde{K}_{\rm g}^{(1)}(k) \ne 0$.

The case, $\widetilde{K}_{\rm g}^{(1)}(k) = 0$ can happen in special
cases. It can happen, for example, when the large-scale linear power
is completely erased by some peculiar form of nonlocal biasing, and
mode-mode coupling from nonlinear scales dominates on linear scales.
It also can happen when the biasing is represented by an even function
of $\dell$, e.g., purely quadratic, $\delta_{\rm g} = \dell^2 -
\langle\dell^2\rangle$, or quartic, $\delta_{\rm g} = \delta_{\rm L}^4
- \langle\delta_{\rm L}^4\rangle$, etc. We assume the galaxy formation
does not have such special form and satisfies the condition
$\widetilde{K}_{\rm g}^{(1)}(k) \ne 0$ in the rest of this paper.

The vanishing stochasticity is an important constraint on large
scales. At this point, the naive introduction of stochasticity $r$ in
linear or quasi-linear regime, as is sometimes done in literatures,
should be cautious. On large scales, $r$ could not be freely adjusted,
but would be close to unity. This fact is already noticed in the paper
by Dekel \& Lahav (1998), in which only a specific, simple model is
considered. The same conclusion is derived by Scherrer \& Weinberg
(1998) based on local galaxy formation scheme. Our conclusions apply
not only to local schemes, but also to nonlocal schemes of galaxy
formation. The difference between local and nonlocal scheme is that
local bias generates constant bias factor, while nonlocal bias
generates scale-dependent one.

If there is a non-unity value of $r$ in large-scale limit, it means
that there is some exotic stochasticity which is not relevant to the
initial density fluctuations, or that $\widetilde{K}_{\rm g}^{(1)}(k)
= 0$. As all the structures in the universe are supposed to be formed
from initial density fluctuations, there is no specific reason to
introduce such kind of exotic processes, at least in the framework of
the present standard theory of structure formation in the universe.

Of course, in nonlinear regime where the approximation of equation
(\ref{eq15}) breaks down, the stochasticity in Fourier space arises by
mode-mode coupling. Since the dynamics of such nonlinear regime is
complex enough to trace analytically, the concept of stochastic bias
is useful especially in this regime. In the nonlinear regime, one
should be aware that the nonlinearity of galaxy-density relation also
dominate and that only parameters $b$ and $r$ are not sufficient to
characterize the biasing properties (Dekel \& Lahav 1998).

The equations (\ref{eq17})--(\ref{eq19}) is valid on scales larger
than a nonlinear scale $2\pi/k_{\rm NL}$, because these equations are
derived from lowest order approximation, which assumes $k^3P_{\rm
L}(k)$ is small. If the nonlocality of galaxy formation is small and
$K_{\rm A}^{(1)}(\bfx)$ is localized on some scales smaller than
$2\pi/k_{\rm g}$, then its Fourier transform, $b_{\rm F}(k)$, will be
constant for $k < k_{\rm g}$. If $k_{\rm g} > k_{\rm NL}$, the bias
factor is constant in the validity region of the equations. This is
equivalent to the purely local galaxy formation. In contrast, if
$k_{\rm g} < k_{\rm NL}$, there appears the scale-dependence of bias
parameter besides mode-mode coupling. This scale-dependence comes from
the nonlocality of the galaxy formation. The $k$-dependence of $b_{\rm
F}(k)$ thus is not negligible on scales below $2\pi/k_{\rm g}$ and its
behavior can be describable by equation (\ref{eq20}) on scales above
$2\pi/k_{\rm NL}$.

Even when there is no stochasticity in Fourier space, it may still
appear in real space on linear scales because of the scale-dependence
of bias parameter, besides mode-mode coupling. Actually, the
galaxy-density cross correlation and galaxy correlation for a smoothed
field with linear smoothing length $R$ are given by
\begin{eqnarray}
   &&
   \langle \delm(\bfx;R) \delg(\bfx;R) \rangle =
   \int \frac{k^2dk}{2\pi^2} P_{\rm mg}(k) W^2(kR) =
   \int \frac{k^2dk}{2\pi^2} b_{\rm F}(k) P_{\rm L}(k) W^2(kR),
\label{eq22}\\
   &&
   \langle \delg(\bfx;R)^{\,2} \rangle =
   \int \frac{k^2dk}{2\pi^2} P_{\rm gg}(k) W^2(kR) = 
   \int \frac{k^2dk}{2\pi^2} b_{\rm F}^{\,2}(k) P_{\rm L}(k) W^2(kR),
\label{eq23}
\end{eqnarray}
where $W$ is a Fourier transform of a smoothing function. This
expression and equation (\ref{eq1}) explicitly show the scale
dependence and stochasticity of the biasing in real space. We define
the following notation for k-space averaging for an arbitrary function
$h(k)$:
\begin{eqnarray}
   \overline{h}(R) = 
   \frac
      {\displaystyle\int\frac{k^2dk}{2\pi^2} h(k) P_{\rm L}(k) W^2(kR)}
      {\displaystyle\int\frac{k^2dk}{2\pi^2} P_{\rm L}(k) W^2(kR)}.
   \label{eq23-1}
\end{eqnarray}
Then, equation (\ref{eq1}) reduces to, simply,
\begin{eqnarray}
   b(R) = \sqrt{\overline{b_{\rm F}^{\,2}}(R)},
   \quad
   r(R) = \frac{\overline{b_{\rm F}}(R)}
             {\sqrt{\overline{b_{\rm F}^{\,2}}(R)}}
   \label{eq23-2}
\end{eqnarray}

To obtain more insight on this scale-dependence and stochasticity, let
us expand the bias parameter $b_{\rm F}(k)$ in terms of $(k/k_{\rm
g})^2$, where $2\pi/k_{\rm g}$ is the scale of nonlocality of galaxy
formation as above (terms of odd power of $k$ do not appear for
reflection symmetry):
\begin{eqnarray}
   b_{\rm F}(k) = 
   b_{\rm F}^{(0)} +
   b_{\rm F}^{(1)} \left(k/k_{\rm g}\right)^2 +
   \frac{b_{\rm F}^{(2)}}{2!}
   \left(k/k_{\rm g}\right)^4 +
   {\cal O} \left(k/k_{\rm g}\right)^6.
\label{eq21}
\end{eqnarray}
With this expansion, equation (\ref{eq23-2}) is expanded in a
straightforward manner, and the results are
\begin{eqnarray}
&&
   b(R) = 
   b_{\rm F}^{(0)} +
   3\gamma^2 b_{\rm F}^{(1)} (R_* k_{\rm g})^{-2}
\nonumber\\&&
   \qquad\qquad +\,   \frac{9\gamma^2}{2b_{\rm F}^{(0)}}
   \left[
      b_{\rm F}^{(0)} b_{\rm F}^{(2)} +
      (1 - \gamma^2) \left(b_{\rm F}^{(1)}\right)^2
   \right]
   (R_* k_{\rm g})^{-4} +
   {\cal O}(R_* k_{\rm g})^{-6},
\label{eq24}
\\&&
   r(R) = 1 -
   \frac92 \gamma^2(1-\gamma^2)
   \left( \frac{b_{\rm F}^{(1)}}{b_{\rm F}^{(0)}} \right)^2
   (R_* k_{\rm g})^{-4} +
   {\cal O}(R_* k_{\rm g})^{-6}.
\label{eq25}
\end{eqnarray}
In this expression, spectral parameters $\gamma$ and
$R_*$ (Bardeen et al.~1986, BBKS, hereafter) are given by
\begin{eqnarray}
   \gamma = 
   \frac{\overline{k^2}(R)}
      {\sqrt{\overline{k^4}(R)}},
   \qquad
   R_* = \sqrt{\frac{3 \overline{k^2}(R)}{\overline{k^4}(R)}}
\label{eq26}
\end{eqnarray}
where notation of the equation (\ref{eq23-1}) is applied. The
parameter $\gamma$ is of order unity, and $R_*$ is of order $R$. {}For
example, if the power spectrum has the form of power-law, $P_{\rm
L}(k) \propto k^n$, and the smoothing function is Gaussian, $W(x)
\propto \exp(-x^2/2)$, these parameters are given by $\gamma^2 =
(n+3)/(n+5)$ and $R_* = (6/(n+5))^{1/2} R$.

Equations (\ref{eq24}) and (\ref{eq25}) give the expression for
stochastic bias parameter from nonlocality of galaxy formation. These
equations represent the minimal stochasticity which is inevitable when
bias is scale dependent. We have neglected the mode coupling near
nonlinear scales. The scale-dependence and stochasticity appears on
the scale of the nonlocality, $2\pi/k_{\rm g}$. On scales larger than
nonlocality, such scale-dependence and stochasticity disappears.
Especially, stochasticity rapidly vanishes for the lack of $R^{-2}$
term in equation (\ref{eq25}). These results do not depend on specific
details of galaxy formation as long as $\widetilde{K}_{\rm g}^{(1)}(k)
\ne 0$, and higher-order terms are negligible. The information of
galaxy formation involves only through two parameters, $b_{\rm
F}^{(0)}$ and $b_{\rm F}^{(1)}$, up to the order $(k_{\rm g} R)^{-2}$,
or three parameters $b_{\rm F}^{(0)}$, $b_{\rm F}^{(1)}$ and $b_{\rm
F}^{(2)}$ up to the order $(k_{\rm g} R)^{-4}$. These parameters are
related to galaxy formation through equation (\ref{eq20}) which should
be calculable if we could know the details of galaxy formation
process. Otherwise, they can be considered as free parameters to be
fitted by observation, instead of fitting functions $b(R)$ and $r(R)$,
which have the infinite degrees of freedom.

\section{SIMPLE MODELS OF NONLOCAL GALAXY FORMATION}
\setcounter{equation}{0}

So far the argument is quite formal. In this section, we consider
specific models of galaxy formation, i.e., the Lagrangian local
biasing models, the cooperative model, and the peak model, as simple
examples. Although the quantitative correspondence of these models and
the actual galaxy formation still needs investigation, these examples
can give qualitative aspects on the nonlocal galaxy formation.

\subsection{Lagrangian Local Biasing Models}

In the Lagrangian local galaxy formation models, the number density of
galaxies is a local function of a smoothed linear density field. The
smoothed density field, however, is a nonlocal function of linear
density field, $\dell$. Thus, in some sense, local galaxy formation
models fall in the category of nonlocal models, with particularly
simple functional form:
\begin{eqnarray}
   \rhog(\bfr) = f(\dells(\bfr)),
   \label{eq40}
\end{eqnarray}
where $f$ is an usual one-variable function of the smoothed linear
density field. The smoothed linear density field, $\dells$, is defined
by
\begin{eqnarray}
   \dells(\bfr) = 
   \int d^3x W^{\rm (real)}_{\rm s} (\bfr - \bfx) \dell(\bfx),
   \label{eq41}
\end{eqnarray}
where $W^{\rm (real)}_{\rm s}$ is a smoothing kernel function with
smoothing length $s$. With this particularly simple form, the bias
parameter in Fourier space, given by equation (\ref{eq20}), reduces to
\begin{eqnarray}
   b_{\rm F} (k) = 
   \frac{\langle f'(\dells)\rangle}{\langle f(\dells)\rangle}
   W_{\rm s}(ks),
   \label{eq42}
\end{eqnarray}
where $W_{\rm s}$ is a Fourier transform of the smoothing kernel
$W^{\rm (real)}_{\rm s}$. All the other higher-order kernels $K^{(n)}$
($n\geq 2$) vanish. This result is essentially the same which is
derived previously in real space (Szalay 1988; Coles 1993):
\begin{eqnarray}
   \xi_{\rm g}(r) =  
   \left(
     \frac{\langle f(\dells) H_1(\dells/\sigls)\rangle}
       {\langle f(\dells)\rangle}
   \right)^2
   \frac{\xi(r)}{\sigls^2},
   \label{eq43}
\end{eqnarray}
since $\langle f(\dells) H_1(\dells/\sigls)\rangle = \sigls \langle
f'(\dells) \rangle$.

In usual nonlocal biasing models, the smoothing length $s$ is taken to
be in nonlinear regime. As we consider linear scales, $W_{\rm s}(ks)
\sim 1$ and there is no scale-dependence on $b_{\rm F}(k)$ and thus
there is no stochasticity in real space. This result is consistent
with the work by Scherrer \& Weinberg (1998), who showed the local
models produce constant bias factor and there is negligible
stochasticity on large scales. Examples of local models are
density-threshold bias, $f(x) \propto \theta(x - \nu \sigls)$, where
$\theta(x)$ is a step function (Kaiser 1984; Jensen \& Szalay 1986),
weighted bias, $f(x) \propto (1 + x) \theta(x - \nu \sigls)$ (Catelan
et al. 1994), Cen-Ostriker bias, $f(x) \propto \exp\{C_0 + C_1 \ln(1 +
x) + C_2 [\ln(1 + x)]^2\}$ (Cen \& Ostriker 1992), etc. In nonlinear
regime, there is really scale-dependence on $b_{\rm F}(k)$ (Mann,
Peacock, \& Heavens 1998; Narayanan, Berlind, \& Weinberg
1998)\footnote{Their work is based on numerical simulation and
Eulerian local biasing, in which the biasing function is a local
function of {\em nonlinear density field}. Thus, strictly speaking,
their analyses are for Lagrangian {\em nonlocal} biasing scheme in the
terminology of this paper. However, the difference between Lagrangian
and Eulerian correlation functions is negligible (higher-order) on
linear scales.}.

Recently, the Mo \& White model (Mo \& White 1996) for the clustering
of dark matter halos, which is an extension of Press-Schechter
formalism (Press \& Schechter 1974; Cole \& Kaiser 1989; Bond et al.
1991), is interested in (e.g., Catelan et al. 1998; Catelan, Matarrese
\& Porciani 1998; Jing 1998; Sheth \& Lemson 1998). This model is also
an example of the Lagrangian local biasing model, because the halo
density contrast $\delta_{\rm h}(1|0)$ in their model is determined
by linear density field $\delta_0$ which is smoothed on scale $R_0$.

\subsection{Cooperative Model}

Next, we consider cooperative galaxy formation model introduced by
Bower et al. (1993), in which they showed that a large-scale ($\sim 20
\himpc$), but weak modulation of galaxy luminosities can reconcile the
discrepancy between the SCDM power spectrum and APM galaxy data (see
also Babul \& White 1991). As a simple example, they identify sites
for galaxy formation as places where density contrast $\delta$
satisfies the following relation
\begin{eqnarray}
   \delta(\bfr) > \nu \sigma - \kappa \bar{\delta}(\bfr),
   \label{eqc01}
\end{eqnarray}
where $\bar{\delta}$ is the density contrast smoothed on a scale
$R_{\rm mod}$, which represent the large-scale modulation, and
$\kappa$ is a constant which is called as ``the modulation
coefficient''. This simple model is mathematically equivalent to the
standard density-threshold bias model, but for the new field defined
by
\begin{eqnarray}
   \delta'(\bfr) = \delta(\bfr) + \kappa \bar{\delta}(\bfr).
   \label{eqc02}
\end{eqnarray}
It is easy to see that this new field is just a new density field with
a smoothing function $W_{\rm s}(ks) + \kappa W_{\rm mod}(k R_{\rm
mod})$, where $W_{\rm s}$ and $W_{\rm mod}$ are smoothing function for
$\delta$ and $\bar{\delta}$, respectively. The bias parameter for the
cooperative model in Fourier space is similar to equation
(\ref{eq42}):
\begin{eqnarray}
   b_{\rm F} (k) &=&
   \frac{\langle d\theta(\delta' - \nu \sigma)/d\delta'\rangle}
      {\langle \theta(\delta' - \nu \sigma)\rangle} 
   \left[
      W_{\rm s}(ks) + \kappa W_{\rm mod}(k R_{\rm mod})
   \right]
   \label{eqc03}\\
   &=&
   \sqrt{\frac{2}{\pi}}
   \frac{e^{-{\nu'}^2/2}}
      {{\rm erfc}\left(\nu'/\sqrt{2}\right)}
   \left[
      W_{\rm s}(ks) + \kappa W_{\rm mod}(k R_{\rm mod})
   \right],
   \label{eqc04}
\end{eqnarray}
where $\nu' = \nu\sigma/\sqrt{\langle {\delta'}^2 \rangle}$. This
expression and equation (\ref{eq23-2}) enable us to evaluate the
stochastic parameters. In Figure 1, we plot the scale dependence of
stochastic parameters of the cooperative model. In this Figure, we
assume a CDM power spectrum of BBKS
\begin{eqnarray}
&&
   P(k) = A k^n 
   \left( \frac{\ln (1 + 2.34 q)}{2.34 q} \right)^2
   \left[
      1 + 3.89 q + (16.1 q)^2 + (5.46 q)^3 + (6.71 q)^4
   \right]^{-1/2},
   \label{eq37}\\
&&
   \qquad\qquad\qquad q = \frac{k}{\Gamma h {\rm Mpc}^{-1}},
\end{eqnarray}
with primordial spectral index $n = 1$ and shape parameter $\Gamma =
0.5$ and linear amplitude $A = 3.35\times 10^5\,(\himpc)^4$ which
corresponds to $\sigma_8 = 1$. The top-hat smoothing function is
applied for smoothing $R$, while the Gaussian smoothing function is
applied for smoothing $s$ and $R_{\rm mod}$:
\begin{eqnarray}
   W(x) = \frac{3}{x^3} (\sin x - x \cos x),
   \qquad
   W_{\rm s}(x) = W_{\rm mod}(x) = e^{- x^2/2},
   \label{eqc05}
\end{eqnarray}
and we set $\nu = 2.8$, $s = 0.5 \himpc$ and $R_{\rm mod} = 10, 20, 30
\himpc$. The modulation coefficient is adjusted so as to produce the
2.5\% rms modulation of the threshold, according to Bower et al.
(1993), i.e., $\kappa \bar{\sigma}/\nu\sigma \equiv \Delta\nu/\nu =
0.025$. This required taking $\kappa = 0.89$, $2.32$, $4.35$ for
$R_{\rm mod} = 10$, $20$, $30\himpc$, respectively\footnote{The small
differences of our value of $\kappa$ and Bower et al.'s are due to 
different fitting formula for power spectrum.}.

As one can see from the Figure, there appears the scale-dependence of
bias parameter $b(R)$ on scales of $R_{\rm mod}$ so that the galaxy
clustering on large scales are enhanced by cooperative bias. This fact
is a main motivation for Bower et al. (1993) to introduce the
cooperative model. The stochastic parameter $r(R)$ is very close to
unity except on the modulation scale, where there is weak
stochasticity due to scale dependence of bias parameter.

\subsection{Peak Model}

Next example of nonlocal galaxy formation is the peak model. In the
peak model, the sites for galaxy formation are identified as high
peaks of initial density field with a fixed smoothing length (see
BBKS). Treating the constraint properly for density peaks is difficult
but there are several approximations. In this paper, we approximate
the density peaks by density extrema (Otto, Politzer \& Wise 1986;
Cline et al. 1987; Catelan et al. 1988). The number density of density
extrema above threshold $\delta_{\rm t}$ is given by
\begin{eqnarray}
   \rhog(\bfr) = 
   \theta(\dells(\bfr) - \delta_{\rm t}) 
   \delta^3\left(\partial_i \dells(\bfr)\right)
   (-1)^3 \det\left(\partial_i \partial_j \dells(\bfr)\right),
   \label{eq28}
\end{eqnarray}
where $\dells(\bfr)$ is a smoothed linear density field with
smoothing length $s$. Density extrema are identical to density peaks
above some moderate threshold where almost all density extrema would
be density peaks.

{}From the general consideration in the previous section, it is
straightforward to obtain parameters $b(R)$ and $r(R)$ of stochastic
bias for this model. From equation (\ref{eq20}) and after tedious
calculation\footnote{For interested readers, we refer Appendix E of
Matsubara (1995) which contains the essential equations to derive the
following equation. One can also show that the higher-order kernels
$\widetilde{K}^{(n)}$ of this model are finite in large-scale
limit.}, we obtain
\begin{eqnarray}
   b_{\rm F}(k) =
   \frac{1}{\sigma_{{\rm s}0} H_2(\nu)}
   \left[
      H_3(\nu) + 
      \frac{s_*^{\,2}}{\gamma_{\rm s}^{\,2}} H_1(\nu) k^2
   \right]
   W_{\rm s}(ks),
\label{eq29}
\end{eqnarray}
where $W_{\rm s}(ks)$ is a Fourier transform of smoothing function for
$\dells$. Other quantities in this expression are defined by
\begin{eqnarray}
&&
   \sigma_{{\rm s}j}^{\,2} =
   \int \frac{k^2dk}{2\pi^2} k^{2j} P_{\rm L}(k)
   W_{\rm s}^{\,2}(ks),
\label{eq30}\\
&&
   \nu = \frac{\delta_{\rm t}}{\sigma_{{\rm s}0}},
   \quad
   s_* = \sqrt{3}\frac{\sigma_{{\rm s}1}}{\sigma_{{\rm s}2}},
   \quad
   \gamma_{\rm s} = 
   \frac{\sigma_{{\rm s}1}^{\,2}}{\sigma_{{\rm s}2}\sigma_{{\rm s}0}}.
\label{eq31}
\end{eqnarray}
Hermite polynomial $H_n$ are defined with the normalization,
\begin{eqnarray}
   H_n(\nu) = e^{\nu^2/2} \left(-\frac{d}{d\nu}\right)^n
      e^{-\nu^2/2}.
\label{eq32}
\end{eqnarray}
Equation (\ref{eq29}) explicitly show the scale-dependence of bias
parameter in Fourier space. The scale of nonlocality corresponds to
$s_*$, which is of order of smoothing length $s$ for obtaining density
peaks.

The stochastic parameters $b(R)$, $r(R)$ in linear regime for density
peaks are derived from the equation (\ref{eq23-2}). Assuming $\nu >
1$, the result is
\begin{eqnarray}
&&
   b(R) = \frac{\nu}{\nu^2 - 1}
   \frac{\sigma_{{\rm c}0}}{\sigma_0 \sigma_{{\rm s}0}}
   \sqrt{
      \left(\nu^2 - 3\right)^2 +
      6 (\nu^2 - 3)
      \frac{\gamma_{\rm c}^{\,2}}{\gamma_{\rm s}^{\,2}}
      \left(\frac{s_*}{c_*}\right)^2 +
      9 \frac{\gamma_{\rm c}^{\,2}}{\gamma_{\rm s}^{\,4}}
      \left(\frac{s_*}{c_*}\right)^4,
   }
\label{eq33}\\
&&
   r(R) = 
   \frac{\sigma_{{\rm d}0}^{\,2}}{\sigma_0 \sigma_{{\rm c}0}}
   \frac{\displaystyle
      \nu^2 - 3 + 
      3 \frac{\gamma_{\rm d}^{\,2}}{\gamma_{\rm s}^{\,2}}
      \left(\frac{s_*}{d_*}\right)^2}
      {\displaystyle
      \sqrt{
         \left(\nu^2 - 3\right)^2 +
         6 (\nu^2 - 3)
         \frac{\gamma_{\rm c}^{\,2}}{\gamma_{\rm s}^{\,2}}
         \left(\frac{s_*}{c_*}\right)^2 +
         9 \frac{\gamma_{\rm c}^{\,2}}{\gamma_{\rm s}^{\,4}}
         \left(\frac{s_*}{c_*}\right)^4}
      },
\label{eq34}
\end{eqnarray}
where spectral indices of various kind are defined as follows:
\begin{eqnarray}
&&
   \sigma_{{\rm c}j}^{\,2} =
   \int \frac{k^2dk}{2\pi^2} k^{2j} P_{\rm L}(k)
   W_{\rm s}^{\,2}(ks) W^2(kR),
   \quad
   c_* = \sqrt{3}\frac{\sigma_{{\rm c}1}}{\sigma_{{\rm c}2}},
   \quad
   \gamma_{\rm c} = 
   \frac{\sigma_{{\rm c}1}^{\,2}}{\sigma_{{\rm c}2}\sigma_{{\rm c}0}},
\label{eq35}\\
&&
   \sigma_{{\rm d}j}^{\,2} =
   \int \frac{k^2dk}{2\pi^2} k^{2j} P_{\rm L}(k)
   W_{\rm s}(ks) W^2(kR),
   \quad
   d_* = \sqrt{3}\frac{\sigma_{{\rm d}1}}{\sigma_{{\rm d}2}},
   \quad
   \gamma_{\rm d} = 
   \frac{\sigma_{{\rm d}1}^{\,2}}{\sigma_{{\rm d}2}\sigma_{{\rm d}0}}.
\label{eq36}
\end{eqnarray}
Equations (\ref{eq33}) and (\ref{eq34}) describe stochastic parameters
in real space.

It is known that if we take both high threshold limit, $\nu
\rightarrow \infty$, and large scale limit, $R \rightarrow \infty$,
the correlation function of the peak model reduces to that of linear
bias, $\xi_{\rm gg} \sim (\nu/\sigma_0)^2 \xi_{\rm mm}$ (Kaiser 1984;
BBKS 1986). This property is easily confirmed from equation
(\ref{eq33}), where $\sigma_{{\rm c}0} =\sigma_{{\rm d}0} =\sigma_0$
for $R \rightarrow \infty$.

In Figure 2, these equations are plotted for $\nu = 3.0$ with $s =1,
5, 10\himpc$, where we assume CDM power spectrum of BBKS, equation
(\ref{eq37}), with primordial spectral index $n = 1$ and shape
parameter $\Gamma = 0.25$ and linear amplitude $A = 1.591\times
10^6\,(\himpc)^4$ which corresponds to $\sigma_8 = 1$. The top-hat
smoothing function is applied for smoothing $R$, while the Gaussian
smoothing function is applied for smoothing $s$:
\begin{eqnarray}
   W(x) = \frac{3}{x^3} (\sin x - x \cos x),
   \qquad
   W_{\rm s}(x) = e^{- x^2/2}.
   \label{eq38}
\end{eqnarray}
The large smoothing lengths, $s = 5, 10 \himpc$ do not correspond to
galaxy formation, but rather they correspond to cluster formation,
since the cluster of galaxies are density peaks of large smoothing
length.

\placefigure{fig2}

As seen in the Figure, once the smoothing scale $R$ exceeds the
smoothing length $s$, which is the scale of galaxy or cluster
formation in this model, the parameter $r$ of stochasticity rapidly
converges to unity, which means there is no stochasticity above that
scale. As stochasticity vanishes, the bias parameter $b$ converges to
a constant on large scales.

\section{DISCUSSION AND CONCLUSIONS}
\setcounter{equation}{0}

In this paper we explicitly derive the stochasticity parameters of the
bias in linear regime from the nonlocality of the galaxy formation. By
using the generalized Wiener-Hermite functionals, we can derive the
two-point correlation on linear scales which is valid even if the
galaxy formation process itself is both nonlinear and nonlocal. This
is in contrast to the usual Taylor expansion which can not treat the
strongly nonlinear features of galaxy formation. Wiener-Hermite
functionals are orthogonal functionals and we do not have to assume
the smallness of $\dell$ itself, and even do not have to know the
exact nonlinear evolution of density contrast, $\delm$. Instead, we
assume only the smallness of correlation function on large scales.

We show that the stochasticity in Fourier space does not exist in
linear regime (except for the constant term which comes from the small
scale behavior of correlation function), and that the biasing
parameter in Fourier space $b_{\rm F}(k)$ is given by
$\widetilde{K}_{\rm g}^{(1)}(k)$. This conclusion is true as long as
the galaxy formation process satisfies the relation,
$\widetilde{K}_{\rm g}^{(1)}(k) \ne 0$, and higher-order kernels
$\widetilde{K}^{(n)}$ ($n\geq 2$) do not increase with scales when
$k\rightarrow 0$. This property in Fourier space is simply because the
galaxy-galaxy and galaxy-mass correlation functions can be expressed
as convolutions of mass correlation function at lowest order of the
expansion by $\xil$ [equation (\ref{eq15})]. A local model of galaxy
formation has a constant bias factor, while the nonlocal model has a
scale-dependent one, besides mode-mode coupling. In the linear regime
where mode-mode coupling is negligible, stochasticity in real space
comes simply from the scale-dependence of the biasing when the galaxy
formation is nonlocal and $\widetilde{K}_{\rm g}^{(1)}(k) \ne 0$.
Thus, naive introduction of stochasticity in Fourier mode in the
linear regime should be avoided. One cannot introduce stochasticity in
Fourier mode in the linear regime simply because of the lack of
knowledge about galaxy formation. If there is any stochasticity in
Fourier mode in the linear regime, it means that there are exotic
process in the galaxy formation which does not come solely from the
initial density field and such process should be correlated on linear
scales, unless galaxy formation process has a special form to satisfy
$\widetilde{K}_{\rm g}^{(1)}(k) = 0$. Such kind of exotic process is
not likely, at least in the present framework of the standard theory
of structure formation in the universe.

We should note our analyses are restricted to the linear regime. In
the nonlinear regime, there are mode coupling from both nonlinearity
of density evolution and nonlinearity of galaxy formation and it makes
the Fourier mode stochastic. In the region where stochasticity is
prominent, the nonlinear density evolution, which is difficult to
track analytically, is also prominent, so that the phenomenological
approach of stochastic bias should be effectively applied in nonlinear
regime (Dekel \& Lahav 1998).

In strongly nonlinear regime, phenomenological approach by the (hyper)
extended perturbation theory (Colombi et al. 1997; Scoccimarro \&
Frieman 1998) can shed light on how nonlinearity makes cross
correlation $r$ deviate from unity. In this theory, the higher order
cumulants of mass density field is given by
$\langle\delta^n\rangle_{\rm c} = S_n \sigma^{2n-2}$, where $S_n$ is a
constant predicted by tree-level perturbation theory, and $\sigma^2 =
\langle\delta^2\rangle$. Although this theory contains an
extrapolation of weakly nonlinear result to strongly nonlinear regime,
it phenomenologically describe the numerical results. In strongly
nonlinear regime, a mere averaging and cumulant are approximately
equivalent in this ansatz: $\langle\delta^n\rangle =
\langle\delta^n\rangle_{\rm c} + {\cal O}(\sigma^{2n-4}) = S_n
\sigma^{2n-2} + {\cal O}(\sigma^{2n-4})$. Thus, if $\delg = \delta^n -
\langle\delta^n\rangle$, one can obtain $\langle\delg^2\rangle =
S_{2n} \sigma^{4n-2} + {\cal O}(\sigma^{4n-4})$ and
$\langle\delg\delta\rangle = S_{n+1} \sigma^{2n} + {\cal
O}(\sigma^{2n-2})$. Finally, one has $r = S_{n+1}/\sqrt{S_{2n}}$. This
value of $r$ depends on the scale, and departs significantly from
unity on small scales.

The conclusion that the stochasticity is weak on linear scales is good
news for determining the redshift distortion parameter $\beta =
\Omega^{0.6}/b$ on linear scales from a redshift survey. The linear
redshift distortion of power spectrum is given by, in the
plane-parallel limit (Kaiser 1987; Pen 1998),
\begin{eqnarray}
   P^{(s)}_{\rm g}(k) =
   \left[
   1 + 2 r(k) \frac{\Omega^{0.6}}{b(k)} \mu^2 + 
   \left(\frac{\Omega^{0.6}}{b(k)}\right)^2 \mu^4
   \right]
   P_{\rm g}(k),
\label{eq39}
\end{eqnarray}
where $P^{(s)}_{\rm g}(k)$ is redshift-space power spectrum of
galaxies, and $\mu$ is a direction cosine of the angle between the
wave vector $k$ and the line of sight [see Hamilton (1992) for an
expression for two-point correlation function and Szalay, Matsubara \&
Landy (1998) for its generalization to non-plane-parallel case]. Since
we see $r(k) = 1$ on linear scales except some special cases, we do
not need to fit $r$ from the observation when we use only the linear
redshift distortion. However, the previous analyses so far usually
assume the bias parameter $b$ as a scale-independent constant. This
assumption is justified if the scale of nonlocality of galaxy
formation is actually below the linear scale. If it is not, the
scale-dependence of $b(k)$ should also be determined by observation
(or by theories, if possible).

The forthcoming large-scale redshift surveys will reveal the galaxy
distribution especially on linear scales, on which we have not had
sufficient data so far. As shown in this paper, the linear clustering
properties are analytically tractable even when the galaxy formation
itself is a too complex phenomenon to analytically track. The
exploration of linear-scale galaxy distribution can overcome our
ignorance of detailed galaxy formation processes, and will give a
great insight on the primordial features of our universe.

\acknowledgements

We thank the anonymous referee for much detailed comments to improve
the original manuscript. This work was supported in part by JSPS
Postdoctoral Fellowships for Research Abroad.

\newpage

\centerline {\bf REFERENCES}
\bigskip

\def\apjpap#1;#2;#3;#4; {\pp#1, {#2}, {#3}, #4}
\def\apjbook#1;#2;#3;#4; {\pp#1, {#2} (#3: #4)}
\def\apjbok#1;#2;#3;#4;#5; {\pp#1, {#2} (#3; #4: #5)}
\def\apjproc#1;#2;#3;#4;#5;#6; {\pp#1, {#2} #3, (#4: #5), #6}
\def\apjppt#1;#2;#3; {\pp#1, #2, #3}

\apjbook Appel,~P., \& de F\'{e}riet,~J.~K. 1926;
Fonctions Hyperg\'{e}om\'{e}triques et Hypersph\'{e}riques,
Polyn\^{o}mes d'Hermite;Paris;Ganthier-Villars;
\apjpap Babul, A., \& White, S. D. M. 1991;MNRAS;253;31{\sc p};
\apjpap Bardeen,~J.~M., Bond,~J.~R., Kaiser,~N. \& Szalay,~A.~S.
1986;ApJ;304;15;
\apjpap Blanton, M., Cen, R., Ostriker, J. P., \& Strauss, M. A.
1998;ApJ;submitted;astro-ph/9807029;
\apjpap Bond, J. R., Cole, S, Efstathiou, G., \& Kaiser, N.
1991;ApJ;379;440;
\apjpap Bower, R. G., Coles, P., Frenk, C. S. \& White, S. D. M.
1993;ApJ;405;403;
\apjpap Catelan, P., Coles, P., Matarrese, S., \& Moscardini, L.
1994;MNRAS;268;966;
\apjpap Catelan, P., Lucchin, F., \& Matarrese, S. 1988;Phys. Rev.
Lett.;61;267;
\apjpap Catelan, P., Lucchin, F., Matarrese, S., \& Porciani, C.
1998;MNRAS;297;692;
\apjpap Catelan, P., Matarrese, S., \& Porciani, C. 1998;ApJ;502;L1;
\apjpap Cen, R. Y., \& Ostriker, J. P. 1992;ApJ;399;L113;
\apjpap Cline, J. M., Politzer, H. D., Rey, S.-J., \& Wise,~M.~B.
1987;Commun. Math. Phys.;112;217;
\apjpap Cole, S., \& Kaiser, N. 1989;MNRAS;237;1127;
\apjpap Coles, P. 1993;MNRAS;262;1065;
\apjpap Colombi, S., Bernardeau, F., Bouchet, F. R., \& Hernquist, L.
1997;MNRAS;287;241; 
\apjpap Davis, M., \& Geller, M. J. 1976;ApJ;208;13;
\apjpap Davis,~M., Efstathiou,~G., Frenk,~C.~S. \& White,~S.~D.~M.
1985;ApJ;292;371; 
\apjpap Dressler, A. 1980;ApJ;236;351;
\apjppt Dekel, A., \& Lahav, O. 1998;preprint;astro-ph/9806193;
\apjpap Fry, J. N. 1984;ApJ;279;499;
\apjpap Fry, J. N. \& Gazta\~naga, E. 1993;ApJ;413;447;
\apjpap Giovanelli, R., Haynes, M. P., \& Chincarini, G. L.
1986;ApJ;300;77; 
\apjpap Guzzo, L., Strauss, M. A., Fisher, K. B., Giovanelli, R., \&
Haynes, M. P. 1997;ApJ;489;37;
\apjpap Hamilton, A.~J.~S. 1992;ApJ;385;L5;
\apjpap Hermit, S., Santiago, B. X., Lahav, O., Strauss, M. A., Davis,
M., Dressler, A., \& Huchra, J. P. 1996;MNRAS;283;709;
\apjpap Jensen, L. G., \& Szalay, A. S. 1986;ApJ;305;L5;
\apjpap Jing, Y.-P. 1998;ApJ;503;L9;
\apjpap Kaiser, N. 1984;ApJ;284;L9;
\apjpap Kaiser, N. 1987;MNRAS;227;1;
\apjpap Loveday, J., Efstathiou, G., Maddox, S. J., \& Peterson, B. A.
1996;ApJ;468;1; 
\apjpap Matsubara, T. 1995;ApJS;101;1;
\apjpap Mann, R. G., Peacock, J. A., \& Heavens, A. F.
1998;MNRAS;293;209;
\apjpap Mo, H. J., \& White, S. D. M. 1996;MNRAS;282;347;
\apjpap Narayanan, V. K., Berlind, A. A., \& Weinberg D. H.
1998;ApJ;submitted;astro-ph/9812002;
\apjpap Otto, S., Politzer, H. D., \& Wise, M. B. 1986;
Phys.~Rev.~Lett.;56;1878;
\apjbook Peebles, P. J. E. 1980;The Large-Scale Structure of the
Universe;Princeton University Press;Princeton;
\apjpap Pen, U. 1998;ApJ;504;601;
\apjpap Press, W. H., \& Schechter, P. 1974;ApJ;187;425;
\apjpap Rees, M. J. \& Ostriker, J. P. 1977;MNRAS;179;541;
\apjpap Santiago, B. X., \& Strauss, M. A. 1992;ApJ;387;9;
\apjpap Scherrer, R. J. \& Weinberg, D. H. 1998;ApJ;504;607;
\apjppt Scoccimarro, R., \& Frieman, J. A. 1998;astro-ph/9811184;;
\apjpap Szalay, A. S. 1988;ApJ;333;21;
\apjpap Szalay, A. S., Matsubara, T. \& Landy, S. D. 1998;ApJ;498;L1;
\apjpap Taruya, A., Koyama, K., \& Soda, J. 1998;ApJ;in
press;astro-ph/9807005; 
\apjpap Taruya, A., \& Soda, J. 1998;ApJ;submitted;astro-ph/9809204;
\apjpap Tegmark, M., \& Blomley, B. C.
1998;ApJL;submitted;astro-ph/9809324; 
\apjpap Tegmark, M., \& Peebles, P. J. E. 1998;ApJ;500;L79;
\apjpap White, S. D. M., \& Frenk, C. S. 1991;ApJ;379;52;

\newpage

\begin{figure}[t]
\begin{center}
   \leavevmode\psfig{figure=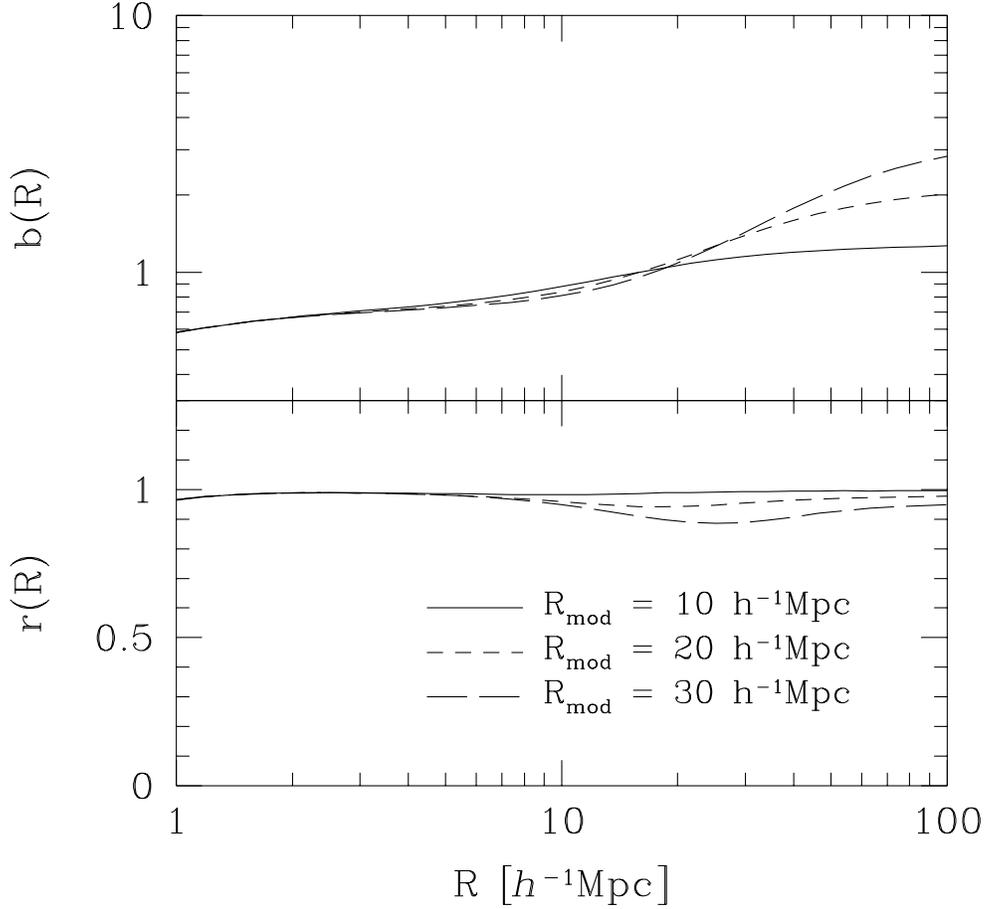,width=16cm,angle=0}
\end{center}
\caption{The stochastic bias parameters $b(R)$ and $r(R)$
for the cooperative model are shown. Underlying density fluctuation is
assumed by CDM model with shape parameter $\Gamma = 0.5$ and amplitude
$\sigma_8 = 1$. Gaussian smoothing scale and the threshold are fixed
as $s = 0.5 \himpc$, $\nu = 2.8$, while the three different modulation
scale $R_{\rm mod} = 10\himpc$ (solid line), $20\himpc$ (dashed line),
$30\himpc$ (long-dashed line) are plotted. The modulation coefficient
is adjusted so as to give a 2.5\% rms modulation of the threshold:
$\kappa = 0.89$, $2.32$, and $4.35$, respectively.
\label{fig1}}
\end{figure}

\begin{figure}[t]
\begin{center}
   \leavevmode\psfig{figure=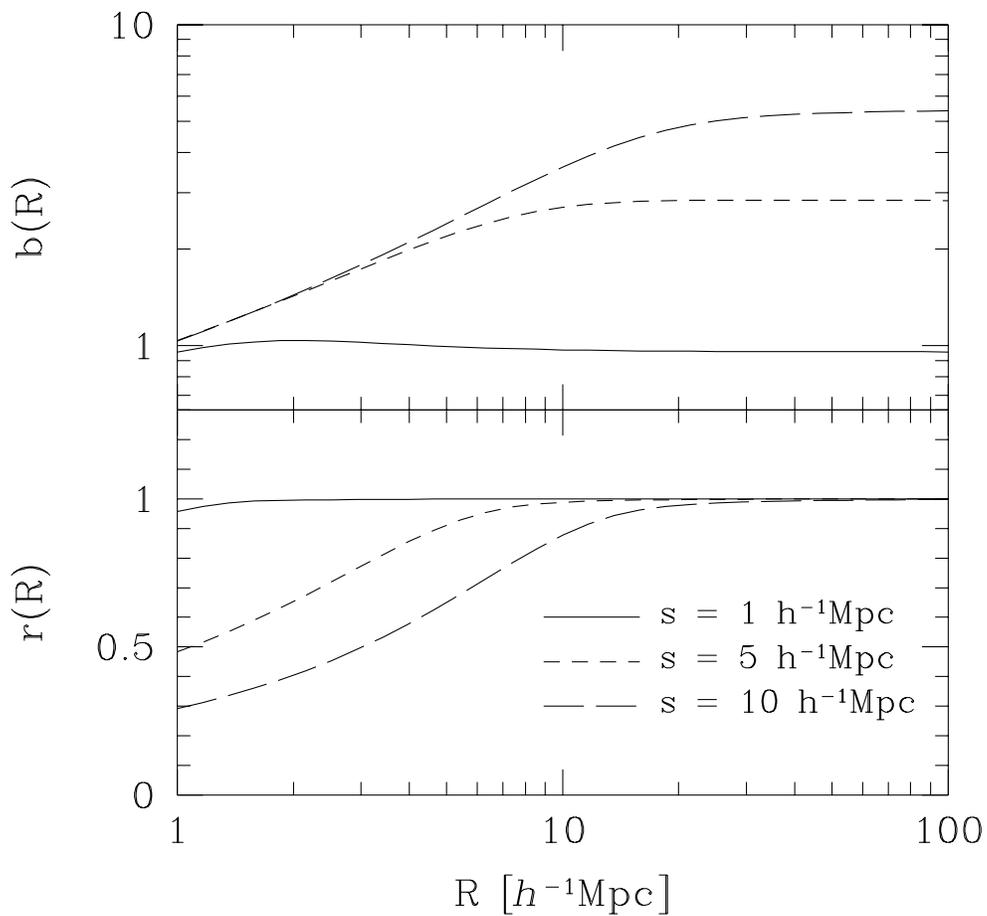,width=16cm,angle=0}
\end{center}
\caption{The stochastic bias parameters $b(R)$ and $r(R)$
for a peak model are shown. Gaussian smoothing scale $s$ which defines
density peaks are varied as $s = 1, 5, 10 \himpc$, which are displayed
by solid lines, short dashed lines, and long dashed lines,
respectively. The threshold for density peaks is fixed as $\nu = 3.0$.
\label{fig2}}
\end{figure}


\end{document}